\documentclass[prb]{revtex4}

\usepackage{graphicx}

\newcommand{\tfrac}[2]{\mbox{$\small\frac{#1}{#2}$}}
\newcommand{\quarterthin}{\kern 0.0417em}

\newcommand{\ket}[1]{|#1\rangle}
\newcommand{\ev}[1]{\langle#1\rangle}

\newcommand{\thin}{\thinspace}

\begin{document}

\title{SU(4) Model of High-Temperature Superconductivity:\\
Manifestation of Dynamical Symmetry in Cuprates}

\author{Yang Sun$^1$, Mike Guidry$^2$, and Cheng-Li Wu$^3$}

\address{
$^1$ Department of Physics, University of Notre Dame, Notre Dame,
Indiana 46556 \\
$^2$ Department of Physics and Astronomy, University of Tennessee,
Knoxville, Tennessee 37996 \\
$^3$ Department of Physics, Chung Yuan Christian University,
Chung-Li, Taiwan 32023 }

\begin{abstract}
The mechanism that leads to high-temperature superconductivity in
cuprates remains an open question despite intense study for nearly
two decades. Here, we introduce an SU(4) model for cuprate systems
having many similarities to dynamical symmetries known to play an
important role in nuclear structure physics and in elementary
particle physics. Analytical solutions in three dynamical symmetry
limits of this model are found: an SO(4) limit associated with
antiferromagnetic order; an SU(2) limit that may be interpreted as
a $d$-wave pairing condensate; and an SO(5) limit that may be
interpreted as a doorway state between the antiferromagnetic order
and the superconducting order. It is demonstrated that with a
slightly broken SO(5) but under constraint of the parent SU(4)
symmetry, the model is capable of describing the rich physics that
is crucial in explaining why cuprate systems that are
antiferromagnetic Mott insulators at half filling become
superconductors through hole doping.
\end{abstract}

\maketitle


\section{Introduction}

Describing collective motion in a strongly correlated many-body
system in terms of single particle degrees of freedom is sometimes
not feasible. Even if one can solve the problem with the help of
large-scale numerical calculations, such a practice may not always
be interesting because physics can be totally buried in the
numerous configurations used in the calculation. In the cuprate
systems exhibiting high-temperature superconducting (SC)
properties, there is a substantial point of view (see, for
example, Ref.\ \cite{An00}) that the many-body correlations are so
strong that the dynamics can no longer be described meaningfully
in terms of individual particles. On the other hand, similar
questions were also asked long ago in nuclear physics. Nuclear
physicists have developed a completely different approach to solve
the many-body problems. The approach is based on the fact that
collective motions in such quantum many-body systems are often
governed by only a few collective degrees of freedom. Once these
degrees of freedom are identified and properly incorporated into a
model, calculations may become feasible and, most importantly, the
physics may become transparent. One systematic way of identifying
the relevant collective degrees of freedom in a many-body system
is the method of dynamical symmetries.

A quantum system exhibiting dynamical symmetries usually contains
two or more competing collective modes. In nuclear physics, it has
long been known that the transition from spherical nuclei, which
dominate the beginnings and endings of nucleonic shells, to
deformed nuclei, which often dominate the middle of shells, is
controlled by the microscopic competition between long-range
quadrupole-quadrupole interactions favoring deformed rotational
mode and short-range monopole pairing interactions favoring
spherical vibrational mode. This competition in nuclear physics
may be expressed algebraically as a competition between a
dynamical symmetry that favors pairing and a dynamical symmetry
that favors multipole (particle--hole) interactions
\cite{FDSM,IBM}. Besides nuclear physics, the dynamical symmetry
method has also been applied in molecular physics
\cite{Mole1,Mole2} and particle physics \cite{Part}.

In cuprates, it is now widely agreed that the mechanism
responsible for superconductivity is closely related to the
unusual antiferromagnetic (AF) insulator properties of their
normal states. There are also compelling arguments that the pair
mechanism leading to high-temperature superconductivity does not
correspond to ordinary BCS $s$-wave pairing. Although experimental
evidence implicates singlet (hole) pairs as the carriers of the
supercurrent, the interaction leading to the formation of the
singlet pairs appears not to be the traditional lattice phonon
mechanism underlying the BCS theory, but rather seems to originate
in strong electron correlations. Furthermore, the pairing gap is
anisotropic, with nodes in the $k_x$--$k_y$ plane strongly
suggestive of $d$-wave hybridization in the 2-particle
wavefunctions.

Such observations argue strongly for a theory based on continuous
symmetries of the dynamical system that is capable of describing
more sophisticated pairing than found in the simple BCS picture
(which is described by a single complex order parameter), and
capable of unifying SC and AF collective modes and the
corresponding phases on an equivalent footing. Then such
fundamentally different physics as SC order and AF order can
emerge from the same effective Hamiltonian as concentration
variables (e.g., doping parameters) are varied. In this
manuscript, we report on an SU(4) model that we have recently
developed \cite{su4a,su4b,su4c,su4d} aiming at understanding the
mechanism that leads to high-temperature superconductivity in
cuprates. Theoretical models employing algebras and groups can be
found in the field of condensed matter (for recent examples, see
Refs. \cite{Zh97,Others}). However, to our knowledge, the powerful
dynamical symmetry method of the type that we shall describe below
has not been applied systematically in this field.

\section{Dynamical Symmetry Method}

A system possesses dynamical symmetries if the Hamiltonian can be
expressed as a polynomial in the Casimir invariants of a subgroup
chain. For approximately the same period of time that the high-Tc
compounds have been known, techniques based on dynamical
symmetries in fermion degrees of freedom have been in development
in the field of nuclear structure physics \cite{FDSM}. There it
has proven fruitful to ask the following questions:  what are the
most important collective degrees of freedom in the low-lying
spectrum of complex nuclei, what are the microscopic many-body
quantum operators that create and annihilate these modes, and what
is the commutator algebra obeyed by this set of operators?

Systematic investigation of these questions has led to strong
confirmation of the following set of conjectures about the nuclear
many-body system:  (1)  Low-lying collective modes are in
approximate one-to-one correspondence with dynamical symmetries in
the fermion degrees of freedom.  (2) A dynamical symmetry
corresponding to low-lying collective modes is associated with a
Lie algebra and its subalgebras that are formed from a set of
fermion operators closed under commutation.  (3) Different
dynamical symmetry subgroup chains arising from the same highest
symmetry group are associated with fundamentally different phases
of the theory. These phases are unified in the highest group. (4)
The unification implied by the preceding point suggests that the
many low-lying collective states formed by systematic filling of
valence shells in heavy nuclei are in reality different
projections in an abstract multidimensional space of the same
state. Equivalently, the different states are transformed into
each other by the generators of the symmetry. Thus, the
systematics of collective modes and phase transitions as a
function of concentration variables are specified by the group
structure.

Motivated by the observation that in a nuclear system at
low-energy excitation, valence nucleon pairs tend to correlate
strongly to form $s$ (spin 0) pairs and $d$ (spin 2) pairs, Wu,
Feng, and Guidry introduced $s$ and $d$ fermion pairs as the basic
building blocks of the Fermion Dynamical Symmetry Model
\cite{FDSM}. These pair operators, when supplemented with the
multipole operators, close either an $SO(8)$ or an $Sp(6)$
algebra, depending on the choice of the basis \cite{FDSM}. If one
neglects the fermionic degrees of freedom of these pairs and
treats them as bosons, one has the early Interacting Boson Model
\cite{IBM} of Arima and Iachello. All the dynamical symmetries
emerging from these models have found correspondence with the
known low-lying collective modes in nuclear systems, such as the
spherical vibrational mode, deformed rotational mode, and
$\gamma$-unstable collective mode.

It has been demonstrated that dynamical symmetries of the type
described in Ref.\ \cite{FDSM} are realized to remarkably high
accuracy in the spectrum and the wavefunctions of microscopic
calculations using the Projected Shell Model \cite{sun98,sun02}.
Since this model is of the shell-model type based on
single-particle degrees of freedom, and is known to give extremely
good agreement with a broad range of experimental data (see, e.g.,
Ref.\ \cite{sun97}), this provides rather conclusive proof that
these dynamical symmetries are strongly realized in the low-lying
states of complex nuclei. This raises the issue of whether similar
symmetries might be found in other complex many-body fermion
systems such as those important in condensed matter.

The dynamical symmetry method applied in cuprates corresponds
schematically to the following algorithm:

0.  Assume the following conjecture:  {\em All strongly collective
modes in fermion (or boson) many-body systems can be put into
correspondence with a closed algebra defining a dynamical symmetry
of the sort described below.}  This is a conjecture, but there is
so much evidence in support of it from various fields of physics
that it is almost a theorem: Strongly correlated motion implies a
symmetry of the dynamics described by a Lie algebra in the
second-quantized operators implementing that motion.

1.  Identify, within a suitable ``valence space'', degrees of
freedom that one believes are physically relevant for the problem
at hand, guided by phenomenology, theory, and general principles.
In the present case, that reduces to defining a minimal set of
operators that might be important to describe superconductivity
and antiferromagnetism.

2.  Try to close a commutation algebra (of manageable dimension)
with the second-quantized operators creating and annihilating the
modes chosen in step 1.  If necessary, approximate these
operators, or add additional ones to the set if the algebra does
not close naturally.

3.  Use standard Lie algebra theory to identify relevant
subalgebra chains that end in algebras for conservation laws that
one expects to be obeyed for the problem at hand. In the present
example, we require all group chains to end in $U(1) \times
SU(2)$, corresponding to an algebra implementing conservation of
charge and spin.

4.  Construct dynamical symmetry Hamiltonians (Hamiltonian that
are polynomials in the Casimir invariants of a group chain) for
each chain. Each such group chain thus defines a wavefunction
basis labeled by the eigenvalues of chain invariants (the Casimirs
and the elements of the Cartan subalgebras), and a Hamiltonian
that is diagonal in that basis (since it is constructed explicitly
from invariants).  Thus, the Schoedinger equation is solved
analytically for each chain, by construction.

5.  Calculate the physical implications of each of these dynamical
symmetries by considering the wavefunctions, spectra, and
transitional matrix elements of physical relevance.  This is
tractable, because the eigenvalues and eigenvectors were obtained
in step 4. Consistency of the symmetry requires that transition
operators be related to group generators; otherwise transitions
would mix irreducible multiplets and break the symmetry.

6.  If step 5 suggests that one is on the right track (meaning
that a wise choice was made in step 1), one can write the most
general Hamiltonian for the system in the model space, which is
just a linear combination of all the Hamiltonians for the symmetry
group chains. Since the Casimir operators of different group
chains do not generally commute with each other, a Casimir
invariant for one group chain may be a symmetry-breaking term for
another group chain.  Thus the competition between different
dynamical symmetries and the corresponding phase transitions can
be studied.

7.  The symmetry-limit solutions may be used as a starting point
for more ambitious calculations that incorporate symmetry
breaking.  Although no longer generally analytical, such more
realistic approximations may be solved by perturbation theory
around the symmetry solutions (which are generally
non-perturbative, so this is perturbation theory around a
non-perturbative minimum), by numerical diagonalization of
symmetry breaking terms, or by coherent state method as shown
later.

Of course, in practical calculations only a few carefully selected
degrees of freedom can be included and the effect of the excluded
space must be incorporated by renormalized interactions in the
truncated space.  It follows that the validity of such an approach
hinges on a wise choice of the collective degrees of freedom and
sufficient phenomenological or theoretical information to specify
the corresponding effective interactions of the truncated space.

\section{The SU(4) Model}

We now introduce a mathematical formalism that provides a
systematic implementation of the dynamical symmetry procedure for
cuprate systems.

\subsection{Choice of operators and the algebra}

The basic assumption of the SU(4) model is that the configuration
space relevant for competing superconductivity and
antiferromagnetism is built from {\it coherent pairs} formed from
two electrons (or holes) centered on adjacent lattice
sites.\footnote{The SU(4) model is particle--hole symmetric. For
cuprate superconductors one is generally interested in hole-doped
compounds but more general applications of SU(4) may be interested
in either electrons or electron holes. Hereafter, unless specified
explicitly, we use ``electrons" to reference either electrons or
holes.} In cuprates the coherent pairs are believed to exhibit
$d$-wave orbital symmetry \cite{Sc95}, suggesting coexistence of
two kinds of coherent pairs in a minimal model: the spin-singlet
($p$) and the spin-triplet ($q$) pairs. We adopt the $d$-wave
pairs with structure defined in Refs.\ \cite{De95,Zh97} as the
basic {\it dynamical building blocks} of the SU(4) model
\begin{equation}
\begin{array}{ll}
\displaystyle p_{12}^\dagger=\sum_k g(k) c_{k\uparrow}^\dagger
c_{-k\downarrow}^\dagger\ \qquad & p_{12} =
(p^\dagger_{12})^\dagger,
\\
\displaystyle q_{ij}^\dagger = \sum_k g(k) c_{k+Q,i}^\dagger
c_{-k,j}^\dagger \qquad & q_{ij} = (q^\dagger_{ij})^\dagger,
\end{array}
\label{D_Pi}
\end{equation}
where $c_{k,i}^\dagger$ creates an electron of momentum $k$ and
spin projection $i,j= 1 {\rm\ or\ }2 \ (\equiv  \uparrow$ or
$\downarrow)$, $g(k)=(\cos k_x-\sin k_y)$ is the $d$-wave form
factor, and $Q=(\pi,\pi,\pi)$ is an AF ordering vector. These pair
operators, when supplemented with operators of particle-hole type
${Q}_{ij}$ and $S_{ij}$,
\begin{equation}
 {Q}_{ij} = \sum_k c_{k+Q,i}^\dagger c_{k,j}
\qquad
 S_{ij} = \sum_k c_{k,i}^\dagger c_{k,j} - \tfrac12 \Omega
\delta_{ij},
\label{Q_S}
\end{equation}
constitute a 16-element  operator set that is closed \cite{su4a}
under a U(4) $\supset$ U(1) $\times$ SU(4) algebra if the
condition
$$
g(k)\approx\mbox{sgn}\,(\cos k_x-\cos k_y)
$$ is imposed.  In Eq.\ (\ref{Q_S}), $\Omega$ is the maximum
number of doped electrons that can form coherent pairs, assuming
the normal state (at half filling) to be the vacuum. The U(1)
factor, ${Q}_+ = {Q}_{11}+{Q}_{22} = \sum_k
(c_{k+Q\uparrow}^\dagger c_{k\uparrow} + c_{k+Q\downarrow}^\dagger
c_{k\downarrow})$, in U(1) $\times$ SU(4) is associated with
charge-density waves, which is independent of the SU(4) algebra.
For no broken pairs the charge-density waves are excluded in the
symmetry limit and in the following discussion we shall restrict
attention to the SU(4) group.

The SU(4) group has three dynamical symmetry group chains
\cite{su4a}:
\begin{eqnarray}
&\supset& {\rm SO(4)} \times {\rm U(1)}
 \supset {\rm SU(2)_s} \times {\rm U(1)} \nonumber \\
{\rm SU(4)} &\supset& {\rm SO(5)} \supset {\rm SU(2)_s} \times
{\rm U(1)} \label{gchain}
\\ &\supset& {\rm SU(2)_p}
\times {\rm SU(2)_s} \supset {\rm SU(2)_s} \times {\rm U(1)}
\nonumber
\end{eqnarray}
each of which ends in the subgroup ${\rm SU(2)_s} \times {\rm
U(1)}$ representing total spin and charge conservation.

The 15 SU(4) generators are related to more physical operators
through the linear combinations:
\begin{equation}
\begin{array}{rcll}
D^\dagger &=& p^\dagger_{12} \qquad D = p_{12} &d-\mbox{wave
singlet pairing (2 operators)}
\\
\vec \pi^\dagger &=& \left(\ i\frac {q_{11}^\dagger\ -
q_{22}^\dagger}2, \ \frac{q_{11}^\dagger + q_{22}^\dagger}2, \
-i\frac {q_{12}^\dagger + q_{21}^\dagger}2 \right) \quad \vec \pi
= (\vec \pi^\dagger)^\dagger \qquad &d-\mbox{wave triplet pairing
(6 operators)}
\\
\vec{\cal Q}\ &=&
\left(\frac{{Q}_{12}+{Q}_{21}}{2},-i\frac{{Q}_{12}-{Q}_{21}}{2},
\frac{{Q}_{11}-{Q}_{22}}{2} \right) &\mbox {staggered
magnetization (3 operators)}
\\
\vec S\ &=& \left( \frac{S_{12}+S_{21}}{2}, \ -i \, \frac
{S_{12}-S_{21}}{2}, \ \frac {S_{11}-S_{22}}{2} \right) &\mbox
{spin (3 operators)}
\\
\hat{n}\ &=& \sum_{k,i} c_{k,i}^\dagger c_{k,i}
=S_{11}+S_{22}+\Omega &\mbox {particle number (1 operator)}
\label{operatorset}
\end{array}
\end{equation}
It will also sometimes prove useful to replace the number operator
$\hat n$ with the charge operator $M$, defined through $
M=\frac12(S_{11}+S_{22})=\tfrac12 (\hat n-\Omega)$.

\subsection{The collective subspace}

The group SU(4) has a quadratic Casimir operator
\begin{equation}
C_{su(4)}=D^\dagger D + \vec \pi^\dagger \cdot \vec \pi + \vec
{\cal Q} \cdot \vec {\cal Q} + \vec S \cdot \vec S + M(M-4).
\label{csu4}
\end{equation}
The group is rank-3 and the irreducible representations (irreps)
may be labeled by 3 weight-space quantum numbers,
$(\sigma_1,\sigma_2,\sigma_3)$. We assume for the simplest
implementation of the model a collective $d$-wave pair subspace
spanned by the following vectors:
\begin{equation}
\ket{S} = \ket{n_x n_y n_z n_s} = (\pi_x^\dagger)^{n_x}
(\pi_y^\dagger)^{n_y} (\pi_z^\dagger)^{n_z} (D^\dagger)^{n_s}
\ket{0}.
\label{collsubspace}
\end{equation}
This collective subspace is associated with irreps of the form
$$
(\sigma_1,\sigma_2,\sigma_3) = (\tfrac \Omega2,0,0),
$$
where 2$\Omega$ is the dimension of the single-particle space in
the SU(4) model, and thus $\Omega$ is the maximum number of pairs
that can be accommodated in the model space. The corresponding
expectation value of the SU(4) Casimir evaluated in these irreps
is a constant,
$$
\ev{C_{su(4)}}=\tfrac\Omega2(\tfrac\Omega2 + 4).
$$

The dimensionality of the full space is $2^{2\Omega}$. However,
the dimensionality of our collective subspace is much smaller,
scaling approximately as $\Omega^4$:
$$
{\rm Dim\thin} (\tfrac{\Omega}{2},0,0) = \tfrac{1}{12}
(\tfrac\Omega 2 + 1)(\tfrac \Omega 2 + 2)^2 (\tfrac \Omega 2 + 3).
$$
Thus for small lattices it is possible to enumerate all states of
the collective subspace in a simple model where observables can be
calculated analytically.

\subsection{SU(4) model Hamiltonian}

The most general 2-body Hamiltonian within the $d$-wave pair space
consists of a linear combination of (quadratic) Casimir operators
$C_g$ for all subgroups $g$
$$
H =H_0+\sum_{g}H_{g}C_{g},
$$
where $H_{g}$ are parameters and the Casimir operators $C_{g}$ are
\footnote{Groups generally may have more than one Casimir
invariant.  We shall use the term ``Casimir'' to refer loosely to
the lowest-order such invariants (which are generally quadratic in
the group generators).  In the context of the present discussion,
quadratic Casimirs are associated with 2-body interactions at the
microscopic level. Higher-order Casimirs are then generally
associated with 3-body and higher interactions. The restriction of
our Hamiltonians to polynomials of order 2 in the Casimirs is then
a physical restriction to consideration of only 1-body and 2-body
interactions.}
\begin{eqnarray}
C_{SO(5)} &=&\vec \pi^\dagger \cdot \vec \pi + \vec S \cdot \vec
S +M(M-3) \nonumber \\
C_{SO(4)} &=& \vec {\cal Q} \cdot \vec {\cal Q} + \vec S
\cdot\vec S \nonumber \\
C_{SU(2)_{p}} &=&D^\dagger D +M(M-1) \label{casimirs}
\\
C_{SU(2)_s} &=& \vec S \cdot \vec S
\nonumber \\
C_{U(1)} &=&M \mbox{ and }M^2. \nonumber
\end{eqnarray}
For fixed electron number the terms in $M$ and $M^2$ in Eq.\
(\ref{casimirs}) are constant. The term $H_0$ is a quadratic
function of particle number and may be parameterized as
$$
H_0 =\varepsilon  n + \tfrac12 \mbox{v}n(n-1),
$$
where $\varepsilon$ and $\mbox{v}$ are the effective
single-electron energy and the average two-body interaction in
zero-order approximation, respectively. Thus the Hamiltonian can
be written as
\begin{eqnarray} H=H_0 + V, \qquad V = - G_0 D^\dagger D -G_1 \vec{\pi}^\dagger \cdot
\vec{\pi} - \chi\vec {\cal Q} \cdot \vec {\cal Q} + \kappa\vec S
\cdot \vec S
\label{ham}
\end{eqnarray}
where $G_0$, $G_1$, $\chi$ and $\kappa$ are the interaction
strengths of $d$-wave singlet pairing, $d$-wave triplet pairing,
staggered magnetization, and spin--spin interactions,
respectively. Since $\ev{C_{su(4)}}$ is a constant, by using Eq.\
(\ref{csu4})  we can eliminate the $\vec \pi^\dagger \cdot \vec
\pi$ term and after renormalizing the interaction strengths the
SU(4) Hamiltonian can be expressed as
\begin{eqnarray}
H = H'_0-G [\ (1-p)D^\dagger D + p\vec {\cal Q} \cdot \vec {\cal
Q}\ ] + \kappa_{\rm eff}\, \vec S \cdot \vec S, \qquad H'_0 =
\varepsilon_{\rm eff}\, n + \tfrac12 \mbox{v}_{\rm eff}\, n(n-1)
\label{generalH} \label{eq20}
\end{eqnarray}
with $(1-p)G=G^{0}_{\rm eff}$, $pG=\chi_{\rm eff}$, and
$\kappa_{\rm eff}$ as the effective interaction strengths, and
where $0\leq p \leq 1$ for the parameter $p$. Since in this paper
we primarily address the ground state properties where $S = 0$,
the last term in Eq.\ (\ref{generalH}) will generally not enter
into the later discussion.

\subsection{Physical interpretation for the dynamical symmetries}

\begin{figure}
  \includegraphics[height=.07\textheight]{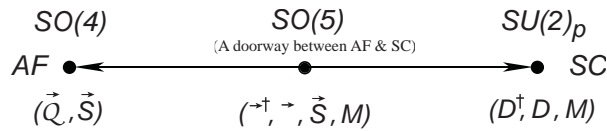}
  \caption{Dynamical symmetries associated with the $U(4)$ symmetry.
The generators are listed for each subgroup.}
\end{figure}

The three subgroup chains of the SU(4) symmetry define three
dynamical symmetries with clear physical meanings. The symmetry
limits SU(2), SO(4), and SO(5), correspond to the choices $p=0$,
1, and 1/2, respectively, in Eq.\ (\ref{eq20}). The Hamiltonian,
eigenfunctions, energy spectrum and the corresponding quantum
numbers of these symmetry limits can be explicitly given (see Ref.
\cite{su4a}). As we shall now discuss, each symmetry limit
represents a different possible low-energy phase of the SU(4)
system. Fig. 1 illustrates these dynamical symmetries and the
relation to the corresponding phases.

The pairing gap $\Delta$ (measure of pairing order) and the
staggered magnetization (measure of AF order) $Q$,
\begin{equation}
\Delta=G^{0}_{\rm eff} \langle D^\dagger D\rangle^{1/2}, \qquad Q=
\langle\vec{\cal Q} \hspace{-2pt}\cdot\hspace{-2pt} \vec{\cal
Q}\rangle^{1/2},
\end{equation}
may be used to characterize the states in these symmetry limits.
We introduce also a relative doping fraction $x$ that is related
to particle number and lattice degeneracy through
\begin{equation}
x=1-\frac{n}{\Omega}.
\label{holedoping}
\end{equation}
Since $\Omega-n$ is the hole number when $\Omega<n$, positive $x$
represents the case of hole-doping, with $x=0$ corresponding to
half filling (no doping) and $x=1$ to maximal doping.

\subsubsection{The SO(4) limit}

The dynamical symmetry chain
$$
 SU(4) \supset SO(4) \times U(1) \supset SU(2)_{s} \times U(1),
$$
which we shall term the SO(4) limit, corresponds to long-range AF
order. This is the symmetry limit of Eq.\ (\ref{eq20}) when $p=1$.
The SO(4) subgroup is locally isomorphic to $SU(2)_F\times
SU(2)_G$, where the product group is generated by the linear
combinations
$$
\vec F= \tfrac12 (\vec {\cal Q} + \vec S), \qquad \vec G =
\tfrac12 (\vec {\cal Q}-\vec S)
$$
of the original SO(4) generators $\vec {\cal Q}$ and $\vec S$. We
may interpret the new generators $\vec F$ and $\vec G$ physically
by noting that if we transform ${\cal Q}_{ij}$ and $S_{ij}$
defined in Eq.\ (\ref{Q_S}) to the coordinate lattice space,
\begin{eqnarray}
        {\cal Q}_{ij}&=&\sum_r (-)^r  c^{\dag}_{ri} c_{rj}
                     = \sum_{r={\rm even}} c^{\dag}_{ri} c_{rj}
                      - \sum_{r={\rm odd}} c^{\dag}_{ri} c_{rj}
\nonumber
\\
              S_{ij}&=&\sum_r c^{\dag}_{ri} c_{rj}
                     = \sum_{r={\rm even}} c^{\dag}_{ri} c_{rj}
                      + \sum_{r={\rm odd}} c^{\dag}_{ri} c_{rj}
\nonumber
\end{eqnarray}
implying that $\vec F$ is the generator of total spin on even
sites and $\vec G$ is the generator of total spin on odd sites.
Thus, we may interpret the SO(4) group as being generated by two
independent spin operators:  one that is the total spin on all
sites and one that is the difference in spins on even and odd
sites of the spatial lattice. This clearly is an algebraic version
of the physical picture associated with AF long-range order.

Furthermore, the ground state has maximal staggered magnetization
$$
Q=\tfrac12 \Omega(1-x)= \tfrac12 n
$$
and a large energy gap (associated with the correlation $\vec{\cal
Q}\hspace{-2pt}\cdot\hspace{-2pt}\vec{\cal Q}$)
$$
\Delta E = 2\chi_{\rm eff}(1-x)\Omega
$$
that inhibits electronic excitation and favors magnetic insulator
properties at half filling. In addition, the pairing gap
$$
\Delta = \tfrac12 G^{0}_{\rm eff} \Omega \sqrt{x(1-x)}
$$
is small near half filling ($x=0$). We thus conclude that these
SO(4) states are identified naturally with an AF insulating phase
of the system.

\subsubsection{The SU(2) limit}

The dynamical symmetry chain
$$
 SU(4) \supset SU(2)_{p} \times SU(2)_{s} \supset
SU(2)_{s} \times U(1),
$$
which we shall term the SU(2) limit, corresponds to SC order and
is the $p=0$ symmetry limit of Eq.\ (\ref{eq20}). In the ground
state, there exists a large pairing gap
$$
\Delta E = G^{(0)}_{\rm eff} \Omega ,
$$
the pairing correlation is the largest among the three symmetry
limits, and the staggered magnetization vanishes in the ground
state:
\begin{equation}
\Delta = \tfrac12 G^{0}_{\rm eff} \Omega \sqrt {1-x^2}\ , \qquad
Q=0. \nonumber\label{DeltaQ}
\end{equation}
Thus we propose that this state is a pair condensate associated
with a $d$-wave SC phase of the cuprates.

\subsubsection{The SO(5) limit}

The dynamical symmetry chain
$$
 SU(4) \supset SO(5) \supset SU(2)_{s} \times U(1),
$$
which we shall term the SO(5) limit, appears when $p=1/2$ in Eq.\
(\ref{eq20}). The SO(5) dynamical symmetry has very unusual
behavior. Although the expectation values of $\Delta$ and $Q$ for
ground state in this symmetry limit are the same as that of Eq.\
(\ref{DeltaQ}) for the SU(2) case, there exists a huge number of
states with different number of $\pi$ pairs that can mix easily
with the ground state when $x$ is small. In particular, at half
filling ($x=0$) the ground state is highly degenerate with respect
to configurations of different number of $\pi$ pairs.  The $\pi$
pairs must be responsible for the antiferromagnetism in this
phase, since within the model space only $\pi$ pairs carry spin.
Thus the ground state in this symmetry limit has large-amplitude
fluctuation in the AF order (and SC order). This indicates that
the SO(5) symmetry limit is associated with phases in which the
system is extremely susceptible to fluctuations between AF and SC
order.

S. C. Zhang proposed to unify AF and SC states by assembling his
order parameters into a five-dimensional vector and constructing
an SO(5) group that rotates AF order into $d$-wave SC order
\cite{Zh97}. The Zhang SO(5) group is a subgroup of our SU(4)
group, implying that the two models are related to each other. The
essential difference is that we implement the full quantum
dynamics (commutator algebra) of these operators exactly, while in
Ref. \cite{Zh97}, the dynamics is implemented in an approximate
manner: a subset of ten of the operators acts as a rotation on the
remaining five operators $\{D^\dagger,D,\vec Q\}$, which are
treated phenomenologically as five independent components of an
order-parameter vector (superspin $\vec n$). Thus only 10 of the
15 generators of our SU(4) are treated dynamically in the Zhang
SO(5) model. If the full quantum dynamics (full commutator
algebra) of the 15 operators is taken into account, the symmetry
for cuprates must be SU(4), not SO(5).

\subsection{Closure of SU(4) and no-double-occupancy constraint}

Suppression of double occupancy on sites in the copper oxide
planes is critical \cite{An87} in explaining why cuprate systems
are antiferromagnetic Mott insulators at half filling and become
superconductors through hole doping. To realize Mott insulator
phases at half-filling it is normal to impose a
no-double-occupancy rule -- the constraint that each lattice site
cannot have more than one valence electron -- by Gutzwiller
projection \cite{SO5proj}. We now show an intimate relationship
between the SU(4) dynamical symmetry and realization of
no-double-occupancy constraint in cuprates.

The relationship can be clearly seen if we express the
momentum-space operator sets (\ref{D_Pi}) and (\ref{Q_S}), with no
approximation to the $d$-wave formfactor $g(k)$, in the coordinate
space as
\begin{eqnarray}
p_{12}^\dagger&=&\sum_{r\in A} \left( c_{{\bf r}\uparrow}^\dagger
c^\dagger_{\bar{\bf r}\downarrow}\ -\ c_{{\bf
r}\downarrow}^\dagger c^\dagger_{\bar{\bf r}\uparrow}\right) \nonumber \\
q_{ij}^\dagger&=&\pm \sum_{r\in A} \left( c_{{\bf r},i}^\dagger
c^\dagger_{\bar{\bf r},j}+c_{{\bf r},j}^\dagger
c^\dagger_{\bar{\bf r},i}\right) \nonumber \\
S_{ij}&=&\sum_{r\in A} \left( c_{{\bf r},i}^\dagger c_{{\bf
r},j}-c_{\bar{\bf r},j} c_{\bar{\bf r},i}^\dagger\right)
\label{rspace2}\\
\tilde{Q}_{ij}&=&\pm \sum_{r\in A} \left( c_{{\bf r},i}^\dagger
c_{{\bf r},j}+c_{\bar{\bf r},j}c_{\bar{\bf r},i}^\dagger\right)
\nonumber
\\
p_{12}&=&(p_{12}^\dagger)^\dagger \qquad
q_{ij}=(q_{ij}^\dagger)^\dagger . \nonumber
\end{eqnarray}
In Eq.\ (\ref{rspace2}), lattice site is separated into $r=$ even
and $r=$ odd sites. The summation is over even (or odd) sites of
the lattice, denoted by $A$, where the $\pm$ sign is + ($-$) if
$A$ is chosen to be even (odd) sites. The quantity
$\tilde{Q}_{ij}$ is defined by $\tilde{Q}_{ij}\equiv {\cal
Q}_{ij}+ \tfrac12 \delta_{ij}\Omega$, where $c^\dagger_{{\bf
r},i}$ ($c_{{\bf r},i}$) creates (annihilates) an electron of spin
$i$ located at ${\bf r}$, while $c^\dagger_{\bar{{\bf r}},i}$
($c_{\bar{{\bf r}},i}$) creates (annihilates) an electron of spin
$i$ at its four neighboring sites, ${\bf r}\pm{\bf a}$ and ${\bf
r}\pm{\bf b}$ with equal probabilities (${\bf a}$ and ${\bf b}$
are the lattice constants along the ${\bf x}$ and ${\bf y}$
directions, respectively, on the copper-oxide plane)
$$
c^\dagger_{\bar{\bf r},i}=\frac 12 \left(c^\dagger_{{\bf r} +{\bf
a},i}+c^\dagger_{{\bf r}-{\bf a},i} -c^\dagger_{{\bf r}+{\bf b},i}
-c^\dagger_{{\bf r}-{\bf b},i}\right) .
$$
Explicit commutation operations show that {\em only if} the
no-double-occupancy constraint is imposed, so that the
anticommutator relation
$$
\left\{c_{\bar{\bf r},i}\ , c^\dagger_{\bar{\bf
r}',i'}\right\}=\delta_{\bf rr'}\delta_{ii'}
$$
is valid, does the operator set (\ref{rspace2}) close under the
${\rm U(1)} \times {\rm SU(4)}$ Lie algebra \cite{su4c}. Thus, the
coordinate-space commutation algebra of the operators
(\ref{rspace2}) demonstrates that SU(4) symmetry {\em necessarily
implies} a no-double-occupancy constraint in the copper oxide
conducting plane. This suggests a fundamental relationship between
SU(4) symmetry and Mott-insulator normal states at half filling
for cuprate superconductors.

One immediate conclusion \cite{su4c} is that the
no-double-occupancy constraint sets an upper limit for the number
of doped holes if SU(4) is to be preserved: $\Omega_{\rm max} =
\tfrac14 \Omega_e$, where $\Omega_e$ is the total number of
lattice sites. Thus, the maximum doping fraction consistent with
SU(4) symmetry is
$$
P_{\rm f}\equiv\frac\Omega{\Omega_e}=\frac 1 4 .
$$
Beyond this doping fraction the no-double-occupancy condition
cannot be ensured and exact SU(4) symmetry cannot be preserved.
The empirical maximum doping fraction 0.23 $\sim$ 0.27 \cite{TS99}
for cuprate superconductivity may then be taken as indirect
evidence for a strongly-realized SU(4) symmetry underlying the
superconductivity in cuprates.

\section{Coherent State Analysis}

The soft nature of the SO(5) phase is seen most clearly if we
introduce the SU(4) coherent states \cite{wmzha90}. The result of
such a coherent state analysis is a set of energy surfaces that
represent an approximation to the original theory, in which order
parameters appear as independent variables. In the general case,
these energy surfaces can exhibit minima and these minima may
appear at non-zero values of the order parameters, implying
spontaneous symmetry breaking.

\subsection{The generalized coherent-state method}

The coherent state method is a well-developed theoretical approach
to relating a many-body algebraic theory with no broken symmetry
to an approximation of that theory that exhibits spontaneously
broken symmetry. This method may be viewed as the extension of
Glauber coherent state theory \cite{Glauber} (which is built on an
SU(2) Lie algebra) to a more complex system having an arbitrary
Lie algebra structure. It has also been shown to be equivalent to
the most general Hatree-Fock-Bogoliubov variational method under
symmetry constraints, and has been applied extensively in various
areas of physics and mathematical physics.

The coherent state $|\psi\rangle$ associated with the SU(4)
symmetry can be written as
\begin{equation}
|\psi\rangle={\cal T}\mid 0^* \rangle,
\label{eq10}
\end{equation}
with the operator ${\cal T}$ defined by
\begin{equation}
{\cal T} = \exp (\eta_{00} p_{12}^{\dagger }+\eta_{10}
q_{12}^{\dagger }-{\rm h.\ c.}).
\label{eq10b}
\end{equation}
In Eq.\ (\ref{eq10}), $|0^{*}\rangle$ is the physical vacuum (the
ground state of the system), the real parameters $\eta_{00}$ and
$\eta_{10}$ are symmetry-constrained variational parameters, and
h.\ c.\ means the Hermitian conjugate. Since the variational
parameters weight the elementary excitation operators
$p^\dagger_{12}$ and $q^\dagger_{12}$, they represent collective
state parameters for a $D$--$\pi$ pair subspace truncated under
the SU(4) symmetry.

It is convenient to use a 4-dimensional matrix representation that
was introduced in Ref.\ \cite{su4b,wmzha88}. The corresponding
matrix elements are defined in the 4-dimensional single-particle
basis
$$
\{ c^\dagger_{{\bf r}\uparrow}|0^*\rangle,\ c^\dagger_{{\bf
r}\downarrow}|0^*\rangle,\ c_{\bar{\bf r}\uparrow}|0^*\rangle,\
c_{\bar{\bf r}\downarrow}|0^*\rangle \}.
$$
The unitary transformation operator ${\cal T}$ of Eq.\
(\ref{eq10}) may be written in this matrix representation as
\begin{equation}
{\cal T} = \left[
\begin{array}{cc} {\bf Y}_1 & {\bf X} \\ -{\bf X}^{\dagger } & {\bf Y}_2
\end{array}
\right] ,   \label{XY-matrix}
\end{equation}
where we define
$$
{\bf X} = \left[
\begin{array}{cc} 0 & v_+\\ -v_- & 0
\end{array}
\right] \quad
 {\bf Y}_1 = \left[
\begin{array}{cc}
u_+ & 0\\ 0 & u_-
\end{array}
\right] \quad {\bf Y}_2 = \left[
\begin{array}{cc}
u_- & 0\\ 0 & u_+
\end{array}
\right],
$$
and where unitarity requires
$$
u_\pm^2+v_\pm^2=1.
$$
The $u$'s and $v$'s, which are related to $\eta_{00}$ and
$\eta_{10}$ in Eq. (\ref{eq10b}), are variational parameters in
the matrix representation that will be interpreted later as
quasiparticle unoccupation and occupation amplitudes,
respectively.

The matrix (\ref{XY-matrix}) implements a quasiparticle
transformation within the $D$--$\pi$ pair space that is further
constrained to preserve the SU(4) symmetry. The physical vacuum
state $|0^*\rangle$ is transformed to a quasiparticle vacuum state
$|\psi\rangle$ and the basic fermion operators,
$$
\{c^\dagger_{{\bf r}\uparrow},c^\dagger_{{\bf r}\downarrow},
c_{\bar{\bf r}\uparrow},c_{\bar{\bf r}\downarrow} \} ,
$$
are converted to quasifermion operators,
$$
\{a^\dagger_{{\bf r}\uparrow}, a^\dagger_{{\bf r}\downarrow},
a_{\bar{{\bf r}}\uparrow}, a_{\bar{{\bf r}}\downarrow} \}
\quad\mbox{with}\quad a_{{\bf r}i}|\psi\rangle=0 ,
$$
through the transformation
\begin{equation} {\cal T}
\left (
\begin{array}{c} c_{{{\bf r}}\uparrow}
\\ c_{{{\bf r}}\downarrow}
\\ c^\dagger_{\bar{\bf r}\uparrow}
\\ c^\dagger_{\bar{\bf r}\downarrow}
\end{array}
\right) \left |0^*\right\rangle= \left(
\begin{array}{c} a_{{{\bf r}}\uparrow}
\\ a_{{{\bf r}}\downarrow}
\\ a^\dagger_{\bar{\bf r}\uparrow}
\\ a^\dagger_{\bar{\bf r}\downarrow}
\end{array}
\right)\left |\psi\right\rangle.
\end{equation}
With above definitions, it is straightforward to write down the
corresponding expectation values of one-body and two-body terms in
Eq.\ (\ref{generalH}) in the coherent-state representation:
\begin{equation}
\begin{array}{rclrcl}
\langle D^{\dagger}\rangle &=&\langle D\rangle=\tfrac12 \Omega
(u_+v_+ +u_-v_-), & \langle D^{\dagger }D\rangle &=& \langle
D^{\dagger}\rangle^2
\\  \langle \pi^{\dagger}_z\rangle &=&\langle \pi_z\rangle=-\tfrac12
\Omega (u_+v_+ -u_-v_-), & \langle \vec{\pi}^{\dagger
}\cdot\vec{\pi}\rangle &=& \langle \pi^{\dagger}_z\rangle^2
\\  \langle {\cal Q}_z\rangle &=&
\tfrac12 \Omega (v_+^2 - v_-^2), & \langle \vec{\cal Q} \cdot
\vec{\cal Q}\rangle &=& \langle {\cal Q}_z\rangle^2
\\
\langle\hat n\rangle &=&\Omega (v_+^2 + v_-^2), & \langle \vec{S}
\cdot \vec{S}\rangle &=& 0
\\
\langle \pi_x\rangle &=& \langle \pi_y\rangle=\langle
\vec{S}\rangle = \langle {\cal Q}_x\rangle=\langle {\cal
Q}_y\rangle=0 &&&
\label{eqone}
\end{array}
\end{equation}
where terms that are of order $1/\Omega$ smaller than the leading
terms have been ignored.

\subsection{Energy surfaces at symmetry limits}

\begin{figure}
  \includegraphics[height=.3\textheight]{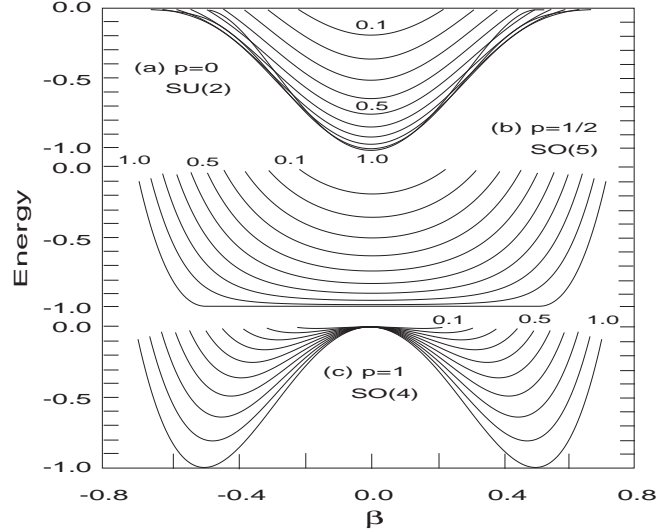}
  \caption{Coherent state energy surfaces.  The energy units are
$G^{(0)}_{\rm{eff}} \Omega^2 / 4$ for figures (a) and (b), and
$\chi_{\rm{eff}} \Omega^2 / 4$ for (c). $H_0$ is taken as the
energy zero point.  Numbers on curves are the effective lattice
occupation fractions, with $n / \Omega = 1$ corresponding to half
filling and $0 < n / \Omega < 1$ to finite hole doping. SO(5)
symmetry corresponds to $p=1/2$ and  the allowed range of $\beta$
is $[-\frac12 \sqrt{n/\Omega},\frac12\sqrt{n/\Omega}]$, which
depends on $n$.}
\end{figure}

With all the expectation values expressed in the coherent state
representation, one can calculate the total energy. In Fig. 2, we
show the coherent-state energy surfaces as a function of the order
parameter $\beta$ for different electron occupation fractions
$n/\Omega$ with $p=0,{1\over 2}$, and 1. The order parameter
$\beta$ is related to the order parameter $Q \equiv \langle {\cal
Q}_z \rangle$ and the electron number $n$ through $\langle {\cal
Q}_z \rangle = 2\Omega\beta\sqrt{n/2\Omega-\beta^2}$. For $p=0$
[SU(2) limit; see Fig. 2(a)], the minimum energy occurs at
$\beta=0$ for all values of $n$, indicating SC order. For $p=1$
[SO(4) limit; see Fig. 2(c)], the opposite situation occurs:
$\beta =0$ is an unstable point and energy minima lie at
$\beta=\pm {1\over 2}\sqrt{n/\Omega}$, indicating the presence of
AF order.

From Fig. 2(b), the SO(5) dynamical symmetry (with $p={1\over 2}$)
is seen to have extremely interesting behavior: the minimum energy
occurs at $\beta=0$ for all values of $n$, as in the SU(2) case,
but there are large-amplitude fluctuations in $\beta$. In
particular, when $n$ is near $\Omega$ (half filling), the system
has an energy surface almost flat for broad ranges of $\beta$.
This suggests a phase very soft against fluctuations in the order
parameters. As $n/\Omega$ decreases, the fluctuations become
smaller and the energy surface tends more and more to the SU(2) SC
limit.

A symmetry limit having such a nature is termed a transitional or
{\em critical dynamical symmetry} \cite{su4b,wmzha88}. A critical
dynamical symmetry is a dynamical symmetry having eigenstates that
vary smoothly with a parameter (usually particle-number related)
such that the eigenstates approximate one phase of the theory on
one end of the parameter range and a different phase of the theory
at the other end of the parameter range, with eigenstates in
between exhibiting large softness against fluctuations in the
order parameters describing the two phases. Thus, we believe that
the rich physics in cuprates, in particular in the underdoped
regime, must lie near the SO(5) limit, subject to a constraint by
the parent SU(4) symmetry.

\subsection{SO(5) symmetry breaking}

Under exact SO(5) symmetry ($p={\frac {1}{2}}$), the AF and SC
states are degenerate at half filling. There is no barrier between
AF and SC states, and one can fluctuate into the other at zero
cost in energy (see the $n/\Omega =1$ curve of Fig.\ 2(b)). This
situation is inconsistent with Mott insulating behavior at
half-filling.  The Zhang SO(5) model has been challenged because
under exact symmetry it does not fully respect the
phenomenological requirements of ``Mottness". As Zhang \cite{Zh97}
has recognized, for antiferromagnetic insulator properties to
exist at half filling, it is necessary to break SO(5) symmetry.
Such breaking of the SO(5) subgroup symmetry is implicit in the
SU(4) model, occurring naturally in the SU(4) model if $p > 1/2$
in the Hamiltonian (\ref{eq20}). Furthermore, the SU(4) symmetry
leads to the following constraint
\begin{equation}
\langle {\cal E}_{su4} \rangle = \langle D^\dagger D+\vec{\cal
Q}\cdot \vec{\cal Q}+\vec{\pi}\cdot\vec{\pi}\rangle = \tfrac14
\Omega^2(1-x^2),
\label{EneSU4}
\end{equation}
which ensures a doping dependence in the solutions. Thus, the
coherent state analysis indicates that the phenomenologically
required SO(5) symmetry breaking and the doping dependence in the
solutions occur naturally in the SU(4) model.

\begin{figure}
  \includegraphics[height=.28\textheight]{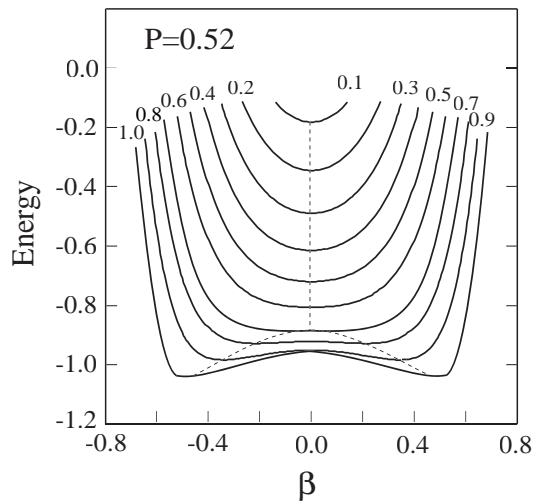}
\caption{Coherent state energy surfaces for slightly perturbed
SO(5). The dotted line indicates the location of the ground state
as $n$ varies and the corresponding $\beta$ value where the energy
curves have a minimum.}
\end{figure}

To see how in the SU(4) model a broken SO(5) symmetry can
interpolate between AF and SC states as particle number varies,
let us perturb slightly away from the SO(5) limit of $p={1\over
2}$. In Fig. 3, SU(4) coherent state results for $p=0.52$ are
shown. One sees from Fig. 3 that if $n$ is near half filling
($n/\Omega=1$), $\beta_{\rm min} = \pm 0.5$; this corresponds to
AF order, since the staggered magnetization reaches its maximum,
$Q=\Omega/2$, and there are no pairing. With the onset of hole
doping, $n/\Omega$ decreases (x increases). The AF correlation $Q$
quickly diminishes and the pairing correlations increase.

The SO(5) symmetry breaking in the Hamiltonian ($p> {1\over 2}$)
is crucial. Only when SO(5) is broken does the energy surface
interpolate between AF and SC order as doping is varied (compare
the surfaces for $p=1/2$  and $p=0.52$ in Figs.\ 2(b) and 3). We
thus conclude that high temperature superconductivity may be
described by a Hamiltonian that conserves SU(4) but breaks
(explicitly) SO(5) symmetry in a direction favoring AF order over
SC order.

\section{Gap diagram at T$=$0}

We have shown that with the coherent state method, one is able to
calculate analytically all the physical quantities at and away
from the symmetry limits. This provides a simple framework to
study in more detail the observables. As our first application, we
use the preceding results to study energy gaps in the cuprate
systems at zero temperature.

\subsection{The gap equations}

We start from the symmetry-constrained variational Hamiltonian
\begin{equation}
  H' = H-\hat{n}\lambda + G {\cal E}_{su4},
\label{1.2}
\end{equation}
where $H$ is the Hamiltonian (\ref{ham}), $\lambda$ and $G$ are
two Lagrange multipliers determined by the constraints of number
conservation, $\langle \hat{n}\rangle=n$, and conservation of the
SU(4) invariant of Eq. (\ref{EneSU4}).

Introducing the energy gaps
\begin{eqnarray}
\Delta_d\equiv G'_0\sqrt{\left\langle D^\dag D\right\rangle}
\qquad \Delta_\pi\equiv G'_1\sqrt{\left\langle\vec{\pi}^\dag\cdot
            \vec{\pi}\right\rangle}
\qquad \Delta_q\equiv\chi'\sqrt{\left\langle\vec{\cal Q}\cdot
            \vec{\cal Q}\right\rangle} ,
\label{1.11}
\end{eqnarray}
where we define
$$
G'_0 \equiv G_0-G \qquad G'_1 \equiv G_1-G \qquad \chi' \equiv
\chi-G,
$$
one obtains
\begin{equation}
\langle  H' \rangle = (\varepsilon-\lambda) n -\left (\
\frac{\Delta_d^2}{G'_0} +\frac{\Delta_\pi^2}{
G'_1}+\frac{\Delta_q^2}{ \chi'}\ \right). \label{hamprime}
\end{equation}
Variation of $\langle H'\rangle$ with respect to $u_\pm$ or
$v_\pm$ (that is, solving $\delta \langle H'\rangle=0$) yields
$$
2u_{\pm}v_{\pm} (\varepsilon_{ \pm}-\lambda)-\Delta_\pm(u^2_{\pm}
-v^2_{\pm})=0,
$$
which is satisfied by
\begin{equation}
\displaystyle u^2_{\pm}=\frac{1}{2}\left
(1+\frac{\varepsilon_{\pm}-\lambda}{e_{\pm}}\right) \qquad
\displaystyle v^2_{\pm}=\frac{1}{2}\left
(1-\frac{\varepsilon_{\pm}-\lambda}{e_{\pm}} \right),
\label{uv}
\end{equation}
with
\begin{eqnarray}
e_{\pm}=\sqrt{(\varepsilon_{\pm}-\lambda)^2+{\Delta_\pm}^2}
\qquad
\Delta_\pm=|\Delta_d\pm\Delta_\pi | \qquad
\varepsilon_{\pm}=\varepsilon\mp\Delta_q .
\label{qse}
\end{eqnarray}
Inserting Eq.\ (\ref{uv}) into Eqs.\ (\ref{eqone}) and
(\ref{EneSU4}), and employing the gap definitions (\ref{1.11}),
one obtains the gap equations
\begin{eqnarray}
\Delta_d&=&\frac{G'_0\Omega}{4}\left ( w_+ \Delta_+  + w_-
\Delta_-\right )
\label{1.24}\\
\Delta_\pi&=&\frac{G'_1\Omega}{4} \left (  w_+ \Delta_+ - w_-\
\Delta_- \right )
\label{1.25}\\
\frac{4\Delta_q}{\chi'\Omega}&=& w_+ ( \Delta_q+\lambda' ) + w_- (
\Delta_q-\lambda' )
\label{1.26}\\
-2x&=&w_+ ( \Delta_q+\lambda' ) - w_- ( \Delta_q-\lambda' )
\label{1.27}\\
\frac{1-x^2}{4}\Omega^2&=& \left(\frac{\Delta_d}{G'_0}\right)^2
+\left(\frac{\Delta_\pi}{G'_1}\right)^2
+\left(\frac{\Delta_q}{\chi'}\right)^2,
\label{1.28}
\end{eqnarray}
where
$$
w_\pm \equiv \frac {1}{e_\pm} , \qquad
     \lambda' \equiv\lambda-\varepsilon\ .
$$
By solving the above algebraic equations, all the gaps and the two
Lagrange multipliers (the chemical potential $\lambda$ and the
SU(4) correlation strength $G$) can be obtained. The three gaps
$\Delta_d$, $\Delta_\pi$ and $\Delta_q$ represent, respectively,
the energy scales of the singlet pairing, triplet pairing, and AF
correlation, and we have introduced an analogous SU(4) gap
$\Delta_{\rm SU4}$ to represent the energy scale of the SU(4)
invariant through
$$
\Delta_{SU4} \equiv G\sqrt{\left \langle{\cal E}_{\rm su4}\right
\rangle}=\tfrac12 G\Omega\sqrt{1-x^2} .
$$
Thus the ground state energy is determined by the four energy gaps
(or energy scales)
\begin{eqnarray}
 E&=&\left \langle H'\right\rangle+n\lambda - \tfrac14 G
(1-x^2)\Omega^2 \nonumber
\\
&=& n\varepsilon -\left (\frac{\Delta_{SU4}^2}{G} +
\frac{\Delta_d^2}{G'_0}
        +\frac{\Delta_\pi^2}{G'_1}
             +\frac{\Delta_q^2}{\chi'} \right ) ,
\nonumber
\end{eqnarray}
where $\left \langle H'\right\rangle$ is given by Eq.
(\ref{hamprime}). Once the gaps and the chemical potential
$\lambda'$ are known, the quasiparticle energies $e_\pm$, and the
amplitudes $u_\pm$ and $v_\pm$, can all be determined through
Eqs.\ (\ref{uv})--(\ref{qse}), permitting other ground state
properties to be calculated.

These results are in many respects analogous to the BCS theory
with $v_{\pm}^2$ the probability of single particle level
$\epsilon_{r\pm}$ being occupied, $\Delta_\pm$ the energy gap, and
$e_{\pm}$ the quasiparticle energies.  The essential difference is
that conventional pairing theories deal with one pairing gap and
one kind of quasiparticles; here we have two pairing energy gaps
and two kinds of interacting quasiparticles, implying a larger
variety of possible behavior. The quantities $e_{\pm}$  are two
kinds of quasiparticle excitation energies, corresponding to two
sets of nondegenerate single particle energy spectra
$\{\varepsilon_{\pm}\}$ separated by $2\Delta_q$. Each level can
be occupied by only one electron of either spin up or spin down.
The corresponding pairing gaps are $\Delta_\pm$, the probabilities
of single-particle levels being unoccupied are $u^2_{\pm}$,  and
the probabilities of single-particle levels being occupied are
$v^2_{\pm}$.

\subsection{Solution of gap equations at T$=$0}

There are three parameters, $\chi$, $G_0$, and $G_1$, in the
coupled algebraic equations (\ref{1.24})--(\ref{1.28}),
corresponding to the three elementary interactions in the SU(4)
model: the AF correlation ($\chi$), the spin-singlet pairing
($G_0$), and the spin-triplet pairing ($G_1$). Physical solutions
of the gap equations depend on the choices for these parameters.
Experimental evidence suggests that these three interactions in
the cuprate system are all attractive, and as demonstrated in Ref.
\cite{su4d}, $\chi$ should be the strongest and $G_1$ the weakest.
Detailed analysis indicates \cite{su4d} that SU(4) symmetric
solutions exist only when $G=G_1$ (that is, when $G'_1 = 0$).
Thus, in the results presented below we shall assume the
conditions
$$
\chi\geq G_0\geq G_1\geq 0\qquad  G = G_1
$$
for the physical coupling constants. By solving Eqs.\
(\ref{1.24}--\ref{1.28}), analytical solutions for the gap
equations at temperature $T=0$ can be obtained. Because of
$G=G_1$, $\Delta_\pi$ is zero for any doping $x$, and the SU(4)
gap is
\begin{equation}
\Delta_{\rm SU4}=\tfrac12 G_1 \Omega \sqrt{1-x^2}.
\label{udodsu4}
\end{equation}

Solution of other gaps at $T=0$ indicates that there is a {\em
critical doping point}
\begin{equation}
x_q=\sqrt{\frac{\chi-G_0}{\chi-G_1}}=\sqrt{1-\frac{G'_0}{\chi'}}
\label{xq}
\end{equation}
separating qualitatively distinct solutions for the quasiparticle
equations. When $x>x_q$, there exists a pure pairing state [all
gaps entering Eqs.\ (\ref{1.24})--(\ref{1.28}) except the singlet
pairing gap $\Delta_d$ vanish]. When $x\leq x_q$, the gap
equations have two solutions: the pairing solution remains but it
is an excited state; another solution becomes the ground state,
which differs from the pure pairing state by having both the AF
gap $\Delta_q$ and the pairing gap $\Delta_d$ nonzero, and
becoming a pure AF state at half filling. Specifically, one finds
that for $x\le x_q$
\begin{eqnarray}
\Delta_q&=&\frac{\chi'\Omega}{2}\sqrt{(x^{-1}_q-
x)(x_q-x)}
\label{udopDq}
\\
\Delta_d&=&\frac{G'_0\Omega}{2}\sqrt{x (x_q^{-1}- x)}
\label{udopDd}
\\
\lambda' &=&-\frac{\chi'\Omega}{2} x_q (1-x_q x),
\label{udopL}
\end{eqnarray}
while for $0<x\leq 1$ (but this is the ground-state solution only
for $x>x_q$),
\begin{eqnarray}
\Delta_q&=& 0
\label{odopDq}\\
\Delta_0 \equiv \Delta_d &=&\frac{G'_0\Omega}{2}\sqrt{(1- x^2)}
\label{odopDd}\\
\lambda' &=&-\frac{G'_0\Omega}{2}x.
\label{odopL}
\end{eqnarray}

\begin{figure}
  \includegraphics[height=.38\textheight]{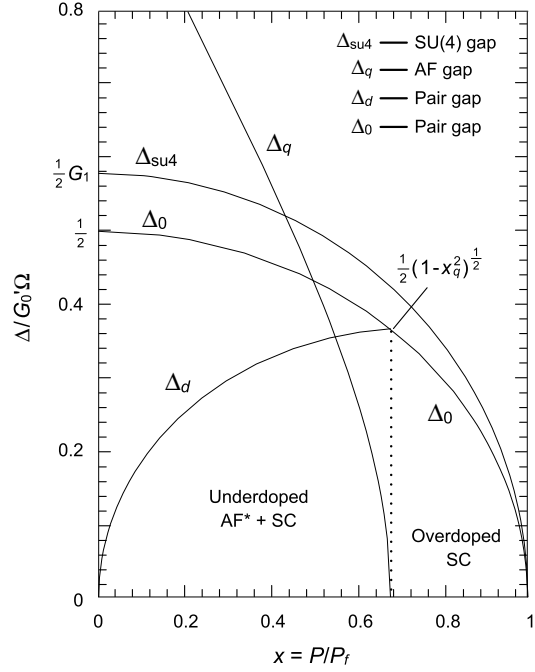}
  \caption{A generic gap diagram for energy gaps versus doping,
as predicted by the SU(4) model. The energy gaps are scaled by the
strength of the pairing correlation $G_0'$ and the doping is
scaled by the maximum doping P$_{\rm f}$. The critical point $x_q$
= 0.68 is chosen in this plot and $\chi' = 17$, $G_0' = 8.2$, and
$G_1 = 9.5$.}
\end{figure}

A generic gap diagram at $T=0$ describing features of the energy
gaps as functions of the doping $x$ is shown in Fig. 4. The
diagram is constructed using the analytical expressions
(\ref{udodsu4}, \ref{udopDq}, \ref{udopDd}, \ref{odopDq},
\ref{odopDd}). It is generic because the relative size of the gaps
can be modified by the choice of interaction strengths but their
basic forms are dictated entirely by the algebraic structure of
the model. Four energy gaps are predicted:
\begin{enumerate}
\item The gap $\Delta_q$ measuring antiferromagnetic correlations
(see Eq. (\ref{udopDq})), which has its maximal value at $x=0$,
decreases nearly linearly to the region of the pairing gaps as
doping increases, crosses the pairing gaps, and vanishes
eventually at the critical doping $x_q$. \item The spin-singlet
pairing gap $\Delta_d$ (see Eq.\ (\ref{udopDd})), which is the
superconducting gap for $x<x_q$. Starting from the critical doping
$x_q$, it decreases as the doping decreases. \item The
spin-singlet pairing gap $\Delta_0= \Delta_d$ (see Eq.\
(\ref{odopDd})), which is the superconducting gap for $x>x_q$ but
is not the order parameter for the ground state in the doping
range $x<x_q$. In $x<x_q$, it increases as the doping decreases.
\item The total SU(4) correlation measured by $\Delta_{\rm SU4}$,
which has its doping dependence following Eq. (\ref{udodsu4}) in
the whole doping range.
\end{enumerate}
It is interesting to observe that at the critical doping value
$x_q$ the pairing gaps exhibit non-trivial behavior: for $x \ge
x_q$, the singlet pairing gap corresponds to a single curve,
labeled by $\Delta_0$. However, as the doping decreases from
$x_q$, the energy scale $\Delta_0$ splits into two curves (labeled
$\Delta_d$ and $\Delta_0$) having very different doping
dependence.

\subsection{Nature of the pseudogaps in cuprates}

The qualitative features of the SU(4) gap diagram in Fig. 4 seem
to agree well with many recent observations in cuprates
\cite{Shen03}. In particular, the appearance of a quantum critical
point $x_q$, the splitting of the pairing gap in the underdoped
regime, and the existence of two distinct pseudogaps are basic
predictions of the SU(4) model that have some experimental
support.

The occurrence of the doping critical point and the splitting of
the pairing gap below the critical point are understood in the
SU(4) model as a direct consequence of the competing pairing and
AF correlations below $x_q$. In this picture, the critical doping
point $x_q$ defines a natural boundary between overdoped and
underdoped regimes. It corresponds to the disappearance of AF
correlations and separates a doping regime characterized by weak
superconductivity and reduced pair condensation energy from a
doping regime characterized by strong superconductivity and
maximal pair condensation energy.

The origin of pseudogaps is one of the unsolved major problems in
high-$T_{\rm c}$ superconductor physics. The pseudogap is believed
to have a profound effect on a wide range of physical properties
of cuprates and is generally considered to be intimately connected
with the origins of high-temperature superconductivity. As noted
in Refs. \cite{Shen03}, there is as yet little consensus about
their origin and behavior but many proposed explanations fall into
one of two categories: the preformed pair picture \cite{EK95} or
the competing order picture \cite{Tallon01}. As we now briefly
discuss (more detail will be published elsewhere \cite{su4d}), the
SU(4) model predicts, and unifies at the microscopic level, two
sources for pseudogap behavior, one of which may be interpreted as
arising from preformed pairs and one of which may be interpreted
as arising from competing order.

The preformed pair picture \cite{EK95} suggests that the pseudogap
is the energy that is needed to form pairs as a first step in
condensing them into a state with long-range order. A pseudogap
arising in this way would be expected to decrease with increased
doping and should merge with the pairing gap in the overdoped
region. The SU(4) gap $\Delta_{\rm SU(4)}$ has exactly these
properties. As can be seen from Eq.\ (\ref{udodsu4}), $\Delta_{\rm
SU(4)}$ depends only on $G_1$ and is independent of $x_q$. It is
maximal at half-filling and falls to zero at the largest doping.
This gap is related to the SU(4) invariant ${\cal E}_{\rm su4}$
(see Eq.\ (\ref{EneSU4})), which is state independent. Therefore,
the SU(4) gap is expected to have no temperature dependence, as
long as the SU(4) symmetry is preserved.

On the other hand, the competing order picture \cite{Tallon01}
suggests that the pseudogap is the energy scale for some form of
order competing with superconductivity. A pseudogap originating in
competing order should not generally merge with the pairing gap
and may fall to zero at a critical doping point where the
competing order is completely suppressed relative to the
superconducting order. The SU(4) antiferromagnetic gap $\Delta_q$
has these properties and gives rise to a pseudogap that may be
interpreted in terms of competing AF and SC order. Specifically,
the scale $\Delta_q$ is the energy per electron required to break
the AF correlation and serves as an order parameter of the AF
phase.

Thus, the SU(4) model predicts the existence of two distinct
pseudogaps. The seemingly rather contradictory mechanisms for the
origin of the pseudogaps as suggested in Refs.
\cite{EK95,Tallon01} may find a consistent explanation in the
SU(4) model.

\section{Conclusion}

The SU(4) symmetry represents a fully microscopic fermion system
in which SC and AF modes enter on an equal footing.  At this
``unification'' level, there is in a sense no distinction among
these degrees of freedom, just as in the Standard Electroweak
Theory of elementary particle physics the electromagnetic and weak
interactions are unified above the intermediate vector boson mass
scale. We may view the system as having condensed into SU(4)
pairs, which fixes the length of the state vectors (through the
SU(4) Casimir expectation value) but not their direction in the
state-vector space. Physically, this means that the system is
paired, with the pair structure exhibiting SU(4) symmetry, but is
neither SC nor AF because fluctuations on a scale set by the
temperature prevent selection of the SC or AF directions. Stated
in another way, the SU(4) pairs are of collective strength, but
are not condensed into a state with long-range order.  Stated in
yet another way, neither the AF nor SC order parameters have
finite expectation values in this regime, but a sum of their
squares (the SU(4) Casimir) does. This constraint implies an
intimate connection between superconductivity and
antiferromagnetism in the SU(4) model. They are, in a sense,
different projections of the same fundamental vector in an
abstract algebraic space.

Compared to the Hubbard or {\em t-J} models, the dynamical
symmetry approach applied here represents a different way to
simplify a strongly-correlated many-body problem. In the Hubbard
or {\em t-J} models, approximations are made to simplify the
Hamiltonian but no specific truncation is assumed for the
configuration space, although in  practical calculations a
truncation is typically necessary.  In contrast, we make no
approximation to the Hamiltonian. The only approximation is the
(severe) space truncation. The symmetry dictated Hamiltonian
includes {\it all} possible interactions in the truncated space.
In principle, if all the degrees of freedom of the system are
included, this approach constitutes  an exact microscopic theory.
The validity of this approach depends entirely on the validity of
the choice of truncated space, which may be tested by comparison
with the data.

The advantage of the dynamical symmetry approach is its cleanness
and simplicity.  It is clean because the {\em only} approximation
is the selection of the truncated model space. Thus, a failure of
the method is a strong signal that one has chosen a poor model
space. It is simple because the method supplies analytical
solutions for various dynamical symmetry limits as a starting
point.  These symmetry-limit solutions provide an immediate handle
on the physics and permit an initial judgement of the model's
validity without large-scale numerical calculations.  Beyond the
symmetry limits, numerical calculations may be necessary.
However, because of the low dimensionality of the models spaces
and the power of group theory, such numerical calculations are
much simpler than those in, say, a Hubbard or {\em t-J} model.

Simple symmetries as a predictor of dynamics has found powerful
application in fields such as particle physics or nuclear physics.
Although a real physical system may be very complex, these
applications are based on the obvious point that nature has
managed to construct a stable ground state having well-defined
collective properties that change systematically from compound to
compound. Extensive experience in many fields of physics suggests
that this is the signal that the phenomenon in question is
described by a {\em small effective subspace with renormalized
interactions} (that may differ substantially in form and strength
from those of the full space) and governed by a symmetry structure
of manageable dimensionality. The results described here suggest
that the observational complexities of high temperature
superconductors may also be described by a simple theory based on
dynamical symmetries having these properties.



\begin{acknowledgments}
We would like to thank Professor V. Zelevinsky and the organizing
committee of {\em Nuclei and Mesoscopic Physics} for including the
present topic into the exciting meeting. Y.S. thanks Professor A.
Aprahamian for her support through the NSF grant under contract
number PHY-0140324.
\end{acknowledgments}


\bibliographystyle{aipproc}   



\bibliographystyle{unsrt}

\end{document}